\newcommand{\be}{\begin{equation}}
\newcommand{\ee}{\end{equation}}
\documentclass[twocolumn,showpacs,superscriptaddress,amsmath,amssymb]{revtex4}
\usepackage{graphicx}% Include figure files
\usepackage{dcolumn}% Align table columns on decimal point
\usepackage{bm}% bold math

\begin{document}
\title{Solitonic lattices in photorefractive crystals\\}

\author{M.~Petrovi\'c}\affiliation{Institute of Physics, P.O.~Box
57, 11001 Belgrade, Serbia}\author{D.
Tr\"ager}\affiliation{Institute of Applied Physics,
Westf\"{a}lische Wilhelms-Universit\"{a}t M\"{u}nster, D-48149
M\"{u}nster, Germany}\author{A.~Strini\'c} \author{M.~Beli\'{c}}\affiliation{Institute
of Physics, P.O.~Box 57, 11001 Belgrade, Serbia}\author{J.
Schr\"oder}\author{C.~Denz} \affiliation{Institute of Applied
Physics, Westf\"{a}lische Wilhelms-Universit\"{a}t M\"{u}nster,
D-48149 M\"{u}nster, Germany}

\date{\today}% any date may be explicitly specified
\begin{abstract}
Two-dimensional spatial solitonic lattices are generated and
investigated experimentally and numerically in an SBN:Ce crystal.
An enhanced stability of these lattices is achieved by exploiting
the anisotropy of coherent soliton interaction, in particular
the relative phase between soliton rows. Manipulation of
individual soliton channels is achieved by use of supplementary control beams.
\end{abstract}
\maketitle Wide ($\sim$1 mm) Gaussian beams launched in a
photorefractive (PR) crystal in the self-focusing regime tend
to break into spatially disordered arrays of filaments, owing to
transverse modulational instabilities \cite{mamaev}. However, ordered arrays
of Gaussian beamlets
($\sim$10 $\mu$m), launched in conditions appropriate to the generation of
spatial screening solitons \cite{opn02}, form much more stable
solitonic lattices. Weakly interacting pixel-like
arrangements of solitons that can individually be addressed are
interesting for applications as self-adaptive waveguides
\cite{petter,chen}.

Adaptive waveguides are of particular interest in all-optical
information processing for their potential to generate large
arrays, as well as for allowing many configurations with different
interconnection possibilities. Spatial optical solitons are
natural candidates for such applications, owing to their ability
for self-adjustable wave\-guiding and versatile interaction
capabilities, as demonstrated in light-induced Y and X couplers,
beam splitters, directional couplers and waveguides. In addition
to such few-beam configurations, the geometries with many solitons
propagating in parallel---the so-called soliton pixels, arrays, or
lattices---have been suggested for applications in information
processing and image reconstruction
\cite{mamyshev,krol,weilnau,bram}. Recently several groups
demonstrated the formation of quadratic arrays of solitons in
parametric amplifiers \cite{bram,minardi} and PR media, with
coherent \cite{weilnau,petter2} and incoherent \cite{chen} beams.

In this communication we combine the properties of spatial PR solitons to
form pixel-like lattices, to
investigate experimentally and numerically the generation and
interactions in large arrays of spatial solitons. We achieve
improved stability of solitonic lattices by utilizing anisotropic
interaction between solitons, in particular the phase-dependent
interaction between solitonic rows. We manipulate individual or
pairs of solitons using incoherent and phase-sensitive control beams.
\begin{figure}%\vspace{5mm}
\includegraphics[width=80mm]{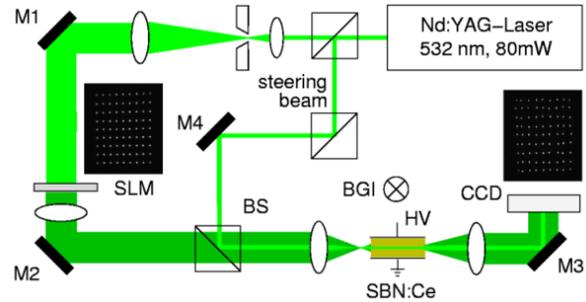}%\vspace{-30mm}
\caption{\label{setup} Experimental setup for the creation of
solitonic lattices and control of individual pixels. Insets:
typical input and output arrays.}
\end{figure}

\begin{figure}%\vspace{-40mm}
\includegraphics[width=80mm]{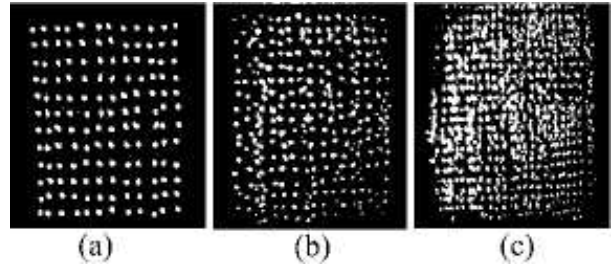}%\vspace{-42mm}
\caption{\label{array-distance} Determining critical separation in a
rectangular lattice, after 20 mm propagation. (a) Above-critical 12 x 12 array with
$\Delta x=70 \mu$m, $\Delta y=85 \mu$m. (b) Slightly below-critical 17 x 17 array with
$\Delta x=60 \mu$m, $\Delta y=75 \mu$m. (c) Below-critical 25 x 25 array with
$\Delta x=50 \mu$m, $\Delta y=65 \mu$m. \vspace{-5mm}}
\end{figure}

\begin{figure}[t]%\vspace{-45mm}
\includegraphics[width=80mm]{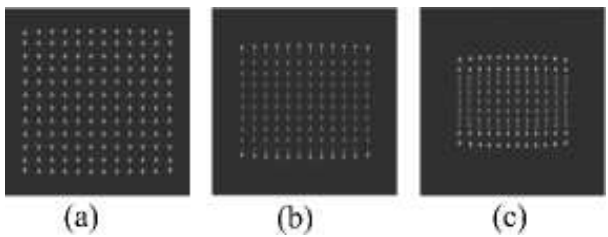}%\vspace{-45mm}
\caption{\label{numlattices} Quadratic
12 x 12 in-phase solitonic lattice after 35 mm propagation, numerical.
Electrooptic coefficient 210 pm/V, external electric field 900
V/cm. The total size of data windows is 1 mm$^2$ and the initial
FWHM size of individual beams 20 $\mu$m. (a) $\Delta x= \Delta
y=70 \mu$m. (b) $\Delta x= \Delta y=60 \mu$m. (c) $\Delta x=
\Delta y=50 \mu$m.\vspace{-7mm}}
\end{figure}

\begin{figure} \vspace{1mm}
\includegraphics[width=80mm]{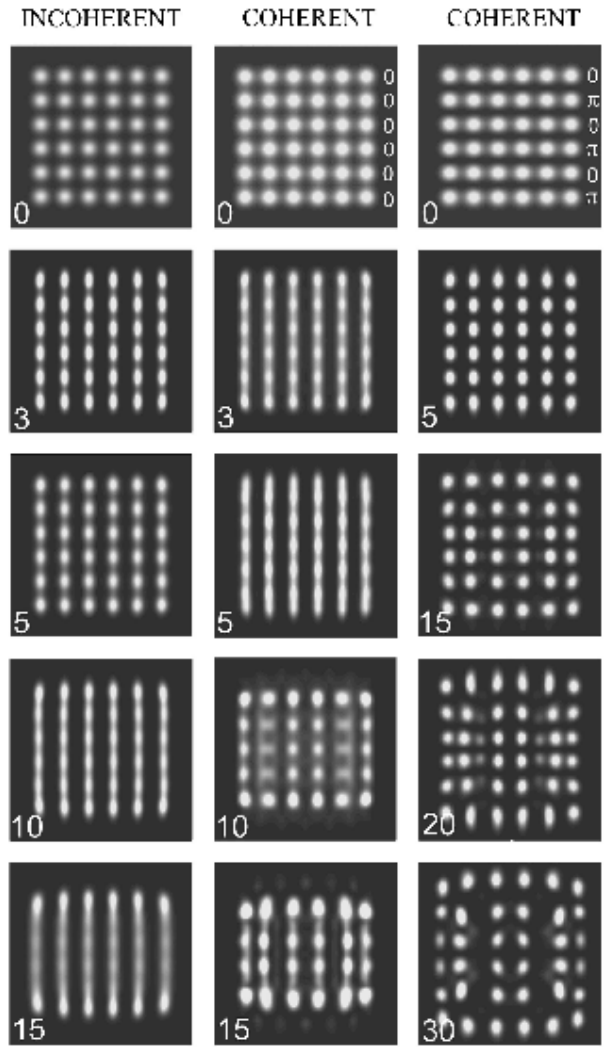}\vspace{8mm}
\caption{\label{slika6big} Comparing the propagation of 6 x 6
sub-critical
square arrays with 40 $\mu$m NN separation, for different
propagation distances, given in each figure. Left column an
incoherent lattice, middle column a coherent in-phase lattice, and
right column a phase-engineered lattice. Other parameters as in
Fig. \ref{numlattices}.\vspace{-4mm}}
%Fig. \ref{numlattices}.}
\end{figure}

\begin{figure}
\includegraphics[width=85mm]{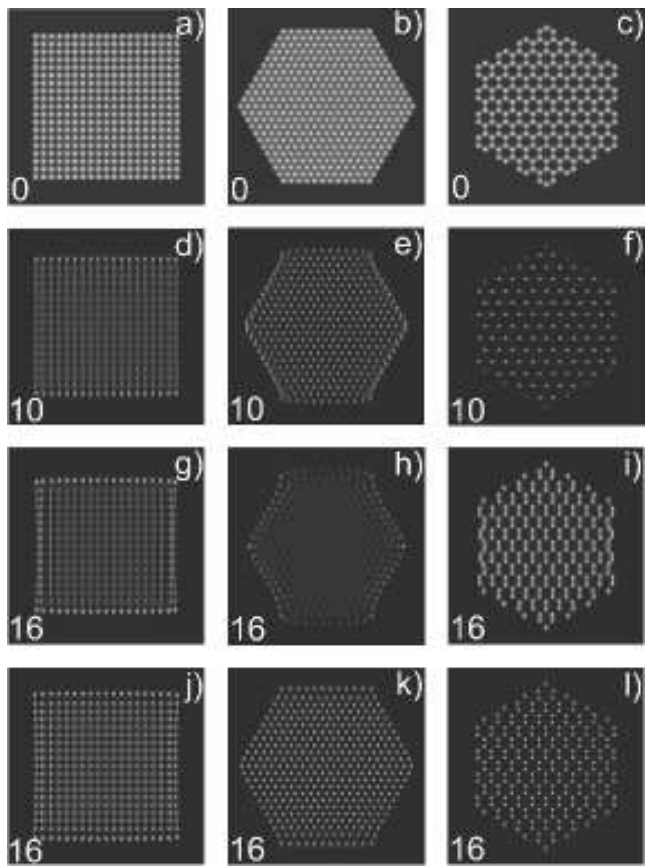}%\vspace{-2mm}
\caption{\label{arrays40x40} Propagation of square, hexagonal, and honeycomb
solitonic lattices with 40 $\mu$m NN separation.
(a)-(c) Initial intensity distributions. (d)-(f)
In-phase lattices propagated for 10 mm. Note in (f) the inversion
of the honeycomb lattice. (g)-(i) In-phase lattices propagated for
16 mm. Strong edge deformation is evident. (j)-(l) The same
lattices, but with alternate rows of solitons out-of-phase,
propagated for 16 mm. Other parameters as in Fig.
\ref{numlattices}.\vspace{-3mm} }
\end{figure}
\begin{figure}%\vspace{-40mm}
\includegraphics[width=80mm]{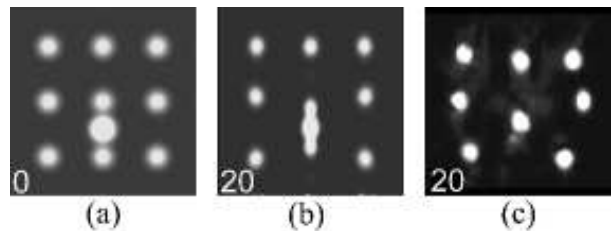}%\vspace{-45mm}
\caption{\label{array-hexagonal} Comparing numerical and
experimental action of a control beam, placed mid-way between two
pixels. (a) Initial configuration. (b) Numerical situation after
20 mm propagation. (c) The corresponding experimental situation.
The data are 70 $\mu$m pixel separation, each spot 20 $\mu$m in
diameter, coherent lattice, incoherent control. The control beam
is three times more intense than any of the pixels.\vspace{-3mm}}
\end{figure}

\begin{figure}
\includegraphics[width=80mm]{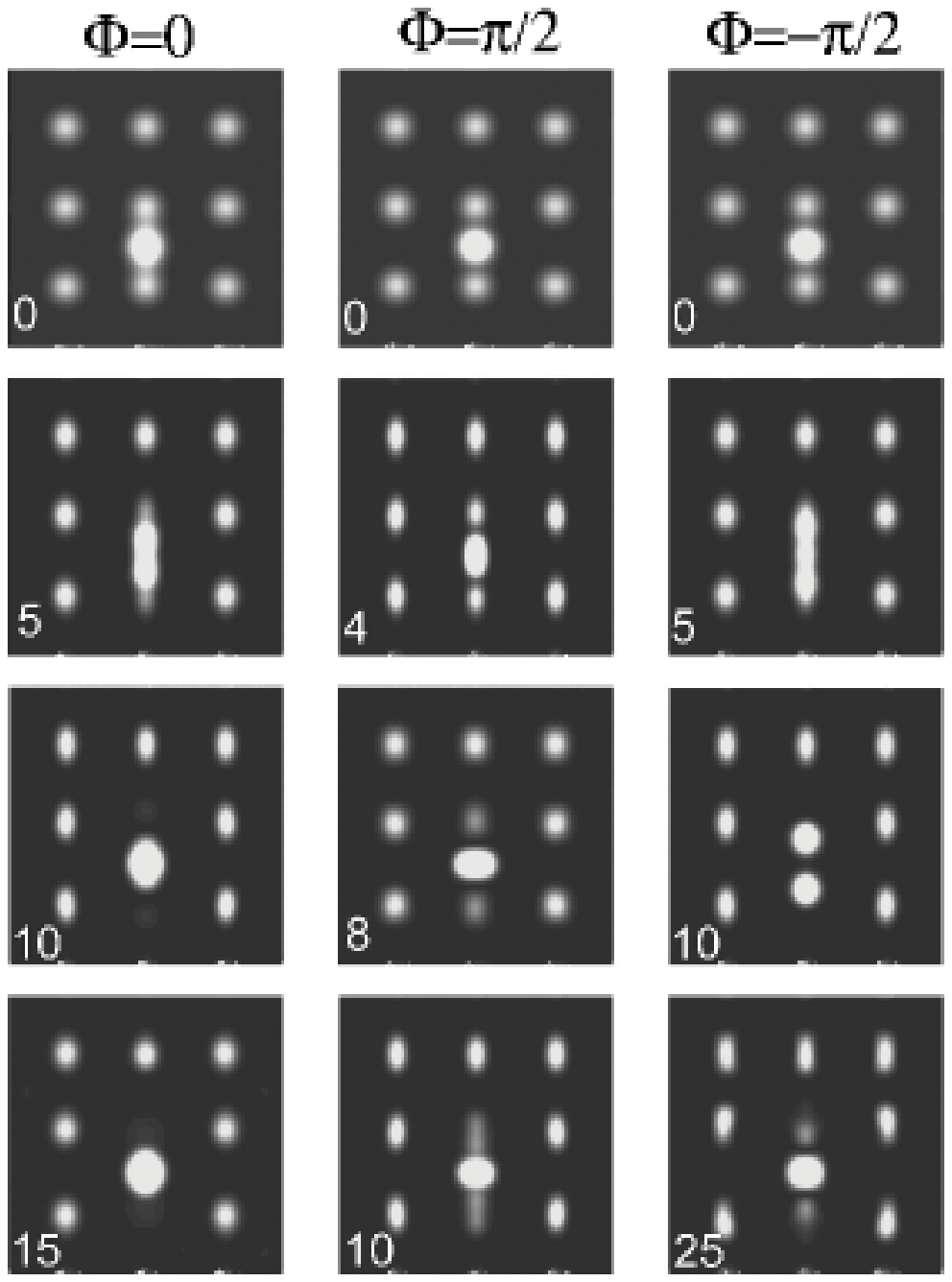}%\vspace{-2mm}
\caption{\label{control} Illustrating the action of a phase-sensitive
control beam in a coherent
square lattice with 70 $\mu$m NN separation. Phase shift of the
control beam relative to the lattice solitons is indicated on the top
of each column. Numbers within each figure
indicate the propagation distance in mm. Other parameters as in
Fig. \ref{numlattices}.\vspace{-4mm}}
\end{figure}

Creation of solitonic lattices requires stable non-interacting
propagation of arrays of self-focusing beams. A crucial feature in
the parallel propagation of PR spatial solitons is their
anisotropic mutual interaction \cite{krol2}. Because the
refractive index modulation induced by a single soliton reaches
beyond its effective waveguide, phase-dependent coherent as well
as separation-dependent incoherent interactions, such as repulsion
or attraction, may appear between the neighboring array elements.
These interactions also affect the waveguiding characteristics of
an individual solitonic channel. Therefore, the separation between
solitons and their nearest-neighbor (NN) arrangement, determined
by the form of the lattice, must carefully be chosen so as to
minimize all forms of interaction. In our experiments, as well as
numerical simulations, we determine the critical soliton distances
for the given crystal thickness at which the interaction becomes
noticeable. To achieve closer packing of solitons, while
maintaining propagation without interaction, the phase
relationship between different beams in the array is exploited.

The creation of lattices and controlling individual pixels
is explored in the experimental setup of Fig. \ref{setup}. Laser
beam derived from a frequency-doubled Nd:YAG laser emitting at 532
nm illuminates a spatial light modulator (SLM), which imprints the
image of a spot-array onto the beam. The spatial light modulator
in turn is imaged onto the front face of a photorefractive
SBN60:Ce crystal (5 x 5 x 20 mm), which is positioned so that the
propagation direction is along the 20 mm axis. To exploit the
dominant electro-optic component $r_{33}\approx 200$ pm/V of our
SBN sample, the incident laser beam is linearly polarized parallel
to the c-axis of the crystal, perpendicular to the propagation
direction. Regular patterns of up to 25 x 25 spots, each with the
diameter of about 20 $\mu$m and an intensity of 110 nW, are imaged
onto the front face of the crystal. Applying an external DC
electric field of about 1 kV/cm across the crystal and
illuminating the crystal with uniform white light to induce an
artificial dark conductivity, creates appropriate conditions for
the formation of spatial screening solitons. To manipulate
individual or pairs of solitons in the lattice, an additional
control beam is derived
from the same laser. It is our experience that large arrays of
solitons can easily be formed in PR crystals. The larger the array
the less significant the finite size (edge) effects. The number of
pixels is mainly limited by the aperture of the PR crystal and the
resolution of the inducing SLM. One can tailor the arrangement of
beams according to specific needs.

Propagation without mutual interaction requires sufficient initial
distance between beams. It also depends on other parameters
influencing the soliton formation and propagation, such as the
initial beam profiles and phases, and the strength of the
nonlinearity, but less crucially. A distance smaller than the
critical distance for coherent interaction between solitons leads
to their attraction and eventual fusion. The attractive
interaction is more pronounced along the y-transverse direction,
perpendicular to the direction of the external field. For a given
propagation distance this is reflected in the deformed appearance
of the lattice. Deformations may also be caused by the crystal
inhomogeneities that can not be controlled. Hence the
determination of the critical distance is qualitative, based on
the inspection of series of runs with decreasing inter-soliton
distances. A typical series of runs is depicted in Fig.
\ref{array-distance}, where the distance between solitons is
reduced by increasing the density of packing. An estimate of the
critical distance is found to lie between 3 and 4 beam diameters
for in-phase arrays in our experimental conditions. It is slightly
larger in the $y$-direction than in the $x$-direction. This estimate
is corroborated in numerical simulations.

Our numerical simulations are based on the paraxial approximation
to the anisotropic propagation of coherent optical beams in
saturable PR media, \be
\label{a2}2ikn_0\partial_z{A}+\nabla^2_\perp{A}= -k^2n_0^4r_{33}
(\partial_x\phi)A\ , \ee where $k$ is the wave number in vacuum,
$n_0\approx2.35$ is the bulk refractive index, $\nabla^2_\perp$ is
the transverse Laplacian, and $A$ the slowly-varying envelope of
the electric field of the beam. The paraxial equation is augmented
with the equation for the electrostatic potential
$-\nabla\phi={\bf E}$ of the space-charge field $\bf E$ generated
in the crystal \cite{zoz,belic1}
\be\nabla^2\phi+\nabla\phi\cdot\nabla\ln(1+I)=E_e\partial_x\ln(1+I)\
,\label{b1}\ee where $E_e$ is the external field. The light
intensity $I$ is normalized to the background intensity. These
equations are solved together, in the manner described in Ref.
\cite{step}. Behavior qualitatively similar to the experimental
was found: Parallel propagation of coherent solitons with varying
separation can be adjusted to be almost without interaction, (Fig.
\ref{numlattices} (a)) whereas for separations below the critical
length the fusion of soliton columns is observed. (Figs.
\ref{numlattices} (b), (c)).

Propagation of incoherent lattices was found to be more stable
than the propagation of coherent in-phase lattices for identical
conditions, however our intention is to exploit the possibility of
manipulating the distribution of phase across the lattice, to
achieve closer packing while maintaining propagation without
interaction. The comparison between incoherent, coherent and
phase-engineered lattice is provided in Fig. \ref{slika6big}. The
NN separation is 40 $\mu$m, which is about half the critical
distance from Fig. \ref{array-distance} and twice the FHWM of
initial beams. It was found that the best results are obtained
when the adjacent rows of solitons are out-of-phase. This is in
agreement with theoretical expectations of the anisotropic
interaction: Along rows the solitons are in-phase, and along
columns they are out-of-phase, which makes the interaction between
them universally repulsive and leads to enhanced stability. It is
seen that the phase-engineered lattice in the right column of Fig.
\ref{slika6big} is much more stable than the incoherent or
coherent in-phase lattices in the left and central columns (note
also different propagation distances). For smaller lattices,
as in Fig. \ref{slika6big}, the edge effects are also more
pronounced.

To further test the idea of phase engineering for enhancing the
stability, different lattices were formed and propagated (Fig.
\ref{arrays40x40}). It is seen that the in-phase lattices deform
fast, and that after 10 mm of propagation, owing to the attractive
forces along the y-direction, some rows are lost in the quadratic
lattice, while the hexagonal and honeycomb lattices are strongly
deformed. The honeycomb lattice deforms continually from the
direct to the inverse (hexagonal) lattice and back (Figs.
\ref{arrays40x40} (f) and (i)). However, out-of-phase lattices for
identical initial conditions and at a longer propagation distance
of 16 mm are only slightly deformed. It is also evident that the
edge effects decrease with increasing the lattice size.

Use of solitonic arrays for applications in information technology
requires means of manipulating individual waveguides, so as to
combine different channels, separate them or induce energy
exchange between the channels. For this purpose the well-known
interaction effects of spatial PR solitons can be exploited
\cite{opn02,weilnau}. We utilize here a supplementary steering
control beam derived from the Nd:YAG laser that can be focused
anywhere on the front face of the crystal. Varying the phase of
the control beam relative to the phase of array elements, a
phase-sensitive coherent or incoherent interaction can be induced,
that may lead to fusion or repulsion of different solitons in the
array. Early experimental results on an incoherent control were
presented in Refs. \cite{petter,chen,weilnau,petter2}. Here we
display the comparison with experiment, using an incoherent
control beam acting on a pair of coherent pixels (Fig.
\ref{array-hexagonal}). The data are as in the experiment of Refs.
\cite{weilnau,petter2}. Qualitative agreement is visible.

The action of a phase-sensitive control beam is depicted in Fig.
\ref{control}. We chose to present the cases of 0 and $\pm\pi/2$
phase shifts of the control beam relative to the lattice. While
for the zero phase shift the merging of the two channels is seen,
in the case of $\pm\pi/2$ a strong energy exchange between the
control beam and the solitons is observed. For $\pi/2$ phase shift
the energy flows from the pixels to the control beam, so that at
approximately 9 mm the control beam is maximally bright and the
pixels are extinguished. For the phase shift of $-\pi/2$ the flow
of energy is from the control beam to the pixels. At approximately
10 mm the control beam is extinguished, and the two slightly
displaced pixels are maximally bright. For the zero phase shift
the three beams gradually merge, so that at approximately 15 mm
only one beam is visible. Afterwards the cycles for the three
values of the phase shift repeat in the inverse order, but less
clearly visible, and some energy is transferred to the adjacent
pixels (not shown).

In summary, we have investigated several aspects of the
interaction of solitons arranged in lattices, for possible use in
all-optical waveguides and interconnects. We have displayed
experimentally the formation of large two-dimensional arrays of
solitons that can be used to guide light at different wavelengths.
Increased stability of solitonic lattices is achieved by the phase
engineering of soliton rows. We have shown how to use a
supplementary steering beam to fuse, extinguish, or enhance
selected channels in the solitonic lattice. Our numerical
simulations, based on the anisotropic model of PR solitons, are in
qualitative agreement with experimental results.\\

AS and MB gratefully acknowledge financial support from the
Alexander von Humboldt Foundation, for the stay and work at WWU
M\"unster. DT acknowledges support from the
Konrad-Adenauer-Stiftung. Parts of this work were supported by the
Graduiertenkolleg "Nichtlineare kontuierliche Systeme" of the
Deutsche Forschungsgemeinschaft. Work at the Institute of Physics
is supported by the Ministry of Science, Technologies, and
Development of the Republic of Serbia, under grants OI 1475 and
1478. We thank Dr. J\"urgen Petter, Institut f\"ur Angewandte
Physik, TU Darmstadt, for help in the experiment on pixel control.

\end{document}